\documentclass[preprint,11pt,5p,twocolumn]{elsarticle}

\usepackage{amssymb}
\usepackage{amsmath}
\usepackage{lineno}
\usepackage{array}
\usepackage{tabularx} 
\usepackage{makecell}
\usepackage{booktabs}
\journal{Acta Materialia}

\begin{document}

\begin{frontmatter}

\title{Deciphering Chemical Ordering in High Entropy Materials: A Machine Learning-Accelerated High-throughput Cluster Expansion Approach}

\author[inst1]{Guillermo Vazquez\corref{cor1}}
\ead{guillermo.vazquez@tamu.edu}
\cortext[cor1]{Corresponding author}

\author[inst1]{Daniel Sauceda}

\author[inst1,inst2,inst3]{Raymundo Arr{\'o}yave\corref{cor1}}
\ead{rarroyave@tamu.edu}

\affiliation[inst1]{organization={Department of Materials Science and Engineering},
            addressline={Texas A\&M University}, 
            city={College Station},
            postcode={77843}, 
            state={TX},
            country={USA}}


\affiliation[inst2]{organization={J. Mike Walker '66 Department of Mechanical Engineering},
            addressline={Texas A\&M University}, 
            city={College Station},
            postcode={77843}, 
            state={Texas},
            country={USA}}

\affiliation[inst3]{organization={Wm Michael Barnes '64 Department of Industrial and Systems Engineering},
            addressline={Texas A\&M University}, 
            city={College Station},
            postcode={77843}, 
            state={Texas},
            country={USA}}

\begin{abstract}
The Cluster Expansion (CE) Method encounters significant computational challenges in multicomponent systems due to the computational expense of generating training data through density functional theory (DFT) calculations. This work aims to refine the cluster and structure selection processes to mitigate these challenges. We introduce a novel method that significantly reduces the computational load associated with the calculation of fitting parameters. This method employs a Graph Neural Network (GNN) model, leveraging the M3GNet network, which is trained using a select subset of DFT calculations at each ionic step. The trained surrogate model excels in predicting the volume and energy of the final structure for a relaxation run. By employing this model, we sample thousands of structures and fit a CE model to the energies of these GNN-relaxed structures. This approach, utilizing a large training dataset, effectively reduces the risk of overfitting, yielding a CE model with a root-mean-square error (RMSE) below 10 meV/atom. We validate our method's effectiveness in two test cases: the (Cr,Hf,Mo,Ta,Ti,Zr)B$_2$ diboride system and the Refractory High-Entropy Alloy (HEA) AlHfNbTaTiZr system. Our findings demonstrate the significant advantages of integrating a GNN model, specifically the M3GNet network, with CE methods for the efficient predictive analysis of chemical ordering in High Entropy Materials. The accelerating capabilities of the hybrid ML-CE approach to investigate the evolution of Short Range Ordering (SRO) in a large number of stoichiometric systems. Finally, we show how it is possible to correlate the strength of chemical ordering to easily accessible alloy parameters. 
\end{abstract}


\end{frontmatter}

\section{Introduction}
\label{sec:intro}

\subsection{Motivation}
Advancements in the aerospace and automotive industries necessitate materials that exhibit exceptional properties---e.g. strength, creep resistance, oxidation resistance, etc.---particularly at elevated temperatures \cite{golla2020review}. Consistent with the objectives of the Materials Genome Initiative, a systematic exploration of the material landscape is imperative to pinpoint materials that meet these rigorous requirements \cite{de2019new}. Beyond the confines of traditional design methodologies, the employment of high-throughput computational strategies and data analytics is indispensable \cite{curtarolo2013high,pollice2021data}. These methods are essential to efficiently traverse the extensive material design space, thereby accelerating the discovery of potentially cutting-edge materials. This paradigm not only optimizes the search process, but also effectively forges a linkage between computational advancements and materials development.

The field of alloy design has been significantly expanded with the introduction of High Entropy Alloys (HEAs) as a concept. Despite the existing nuances among different classes of HEAs, they are distinguished primarily for their chemical complexity and their ability to form stable single-phase solid solutions, as a result of the high entropy of mixing at nearly equiatomic concentrations \cite{tsai2014high,miracle2017critical}. This paradigm shift towards HEAs introduces a novel approach to alloy property engineering through compositional tuning. However, the vastness of the HEA space poses a formidable challenge, as the number of potential alloys far exceeds the capabilities of exhaustive experimental and computational searches. To navigate this extensive material domain effectively, material scientists are increasingly turning to sophisticated frameworks that synergize computational simulations with data engineering \cite{borg2020expanded,senkov2015accelerated}. Such modern approaches are crucial for the efficient exploration of the HEA landscape, facilitating the discovery and refinement of materials that are precisely engineered to meet the specific requirements of diverse industries.

\subsection{Simulating High Entropy Materials}
Much of the current work is focused on strictly equiatomic concentrations of elements. The early transition metal High Entropy Alloys (HEAs) subspace is an extensively researched subclass within the broader HEA family, particularly valued for applications that require sustained performance at elevated temperatures. Research in this field predominantly focuses on the formation of single-phase solutions, specifically Body-Centered Cubic (BCC) and Face-Centered Cubic (FCC) structures. The investigation of the phase stability in these alloy spaces can generally be carried out within the formalism of alloy theory and, more specifically, within the cluster expansion formalism. For example, Nataraj et al. \cite{nataraj2021systematic,nataraj2021temperature} investigated the thermodynamics and phase stability in the  NbTiVZr, HfNbTaTiZr, and AlHfNbTaTiZr alloy systems---important in the context of high-temperature applications \cite{lin2015effect, senkov2014effect}---using a combination of Density Functional Theory (DFT) and Cluster Expansion (CE). 

The CE method  \cite{sanchez1984generalized,zunger1994first,de1994cluster} efficiently approximates the total energy of any configuration in a substitutional alloy by considering cluster contributions. Specifically, it computes the energy of a given configuration within a system that maintains a consistent parent lattice, which is defined by the symmetry of the clusters. This approach not only simplifies the representation of complex alloy configurations but also allows for a more manageable exploration of the vast compositional landscape of single-solution alloys.

The fundamental idea behind cluster expansion is to express the energy of a system as a series of interactions among clusters of atoms. These clusters can range from single atoms to pairs, triplets, and higher-order groupings. The energy of the entire system is then approximated as a sum of these interactions, with each term in the sum corresponding to a specific type of cluster. The coefficients of these terms are determined through fitting to data---i.e. energies of atomic arrangements--- typically obtained from first-principles calculations, such as those from density functional theory (DFT) \cite{kresse1993ab,kresse1994ab,kresse1996efficiency,kresse1996efficient}. 

The Cluster Expansion (CE) method excels in analyzing disordered materials, offering a more comprehensive approach than Density Functional Theory (DFT) to the study of alloying behavior. Yet, its high effectiveness comes with a significant drawback: the complexity and computational demand increase substantially with more complex systems. This complexity arises primarily from the need for extensive data to accurately train the CE surrogate model. As we add more elements to a system, the number of required clusters for a precise representation grows exponentially, leading to increased computational costs \cite{kormann2017long}. In the work by Nataraj et al. \cite{nataraj2021systematic,nataraj2021temperature}, for example, converged CE for the NbTiVZr, HfNbTaTiZr, and AlHfNbTaTiZr alloy systems required
2,984, 1,970, and 4,000 structures respectively. While these calculations are technically feasible, they represent a far from trivial endeavor---each converged DFT calculation takes an order of 100-1,000 core-hours to complete in a typical high-performance research computing facility---and must be replicated for any compositional modification. 

This issue is especially pronounced in the context of High Entropy Materials (HEMs). Implementing CE for HEMs is challenging and infrequent, primarily due to the high costs involved in developing a dependable model \cite{zhu2023probing, nataraj2021temperature, fernandez2017short}. As a result, researchers often lean towards other methodologies \cite{wei1990electronic, zunger1990special, yonezawa1973coherent, faulkner1980calculating}.  Special Quasirandom Structures (SQS) \cite{zunger1990special} are used to construct periodic analogs to random alloys through the use of specially constructed unit cells that mimic (up to a certain coordination range) the statistics of random configurations. The Coherent Potential Approximation (CPA) \cite{yonezawa1973coherent,vela2023high} calculates the electronic properties of disordered systems under an "effective medium" approximation. These approaches demand fewer computational resources at the expense of making some unwarranted assumptions and approximations. 

However, It is worth noting that in the limited cases where CE has been applied to systems with four or more elements, it has proven to be exceptionally effective. Not only does it accurately predict the thermodynamic behavior of these systems, but it also helps understand their stability across various compositions \cite{zhu2023probing, nataraj2021systematic}. Additionally, the CE formalism enables \cite{nataraj2021temperature} the direct assessment of the evolution of Short Range Order (SRO)---this is important as SRO has been shown to have important effects on the properties of HEAs \cite{alvarado2023predicting}. Despite the heavy computational requirements, CE's unrivaled ability to analyze complex multicomponent systems like HEAs makes it a powerful tool for the analysis of the stability and behavior of High Entropy Materials.

\subsection{Proposed Approach}
A good quality CE for a disordered material relies on high-quality DFT calculations. DFT methods, while powerful, are computationally expensive and can significantly constrain the cluster expansion (CE) approach due to the extensive resources required to generate data for fitting the CE coefficients. In response to this challenge, our framework incorporates the M3GNet network \cite{chen2022universal}, a robust universal machine learning (ML) model designed for predicting fundamental properties of crystal structures. M3GNet is first deployed as a zero-shot model to approximate the energetic of various configurations on a lattice, configurations that are traditionally evaluated using DFT methods. Initially, energetic properties predicted by M3GNet are validated with results from conventional DFT calculations. These initial comparisons are then used to retrain the M3GNet model, thereby enhancing its predictive accuracy by fine-tuning its parameters through high-quality DFT data relevant to the system under investigation. Once retrained, M3GNet acts as an efficient surrogate for exhaustive DFT calculations, allowing for the rapid and precise estimation of energetics in complex configurations, particularly those involving a large number of components. This innovative approach markedly accelerates the exploration process, presenting a viable and efficient alternative to the reliance on laborious DFT calculations for extensive and intricate evaluations. To validate the effectiveness of our framework, we explore the alloying behavior of two refractory systems: the (Cr,Hf,Mo,Ta,Ti,Zr)B$_2$ diboride system and the High-Entropy Alloy (HEA) AlHfNbTaTiZr system.

Refractory High Entropy Alloys (RHEAs) \cite{ouyang2023design,vela2023high} are a further narrowed subset of HEAs largely composed of "refactory" constituent elements. There is growing interest in these alloys for their potential application in technologies subject to extreme conditions. These alloys are typically BCC-based solid solutions with very high strengths at elevated temperatures and have been focus of a growing number of publications over the past  decade. The AlHfNbTaTiZr is an exemplar of this alloy space and has been shown to exhibit high strength \cite{lin2015effect}. Compared to earlier RHEAs like TaNbWMo and TaNbWMoV, these alloys exhibit significantly lower density and enhanced ductility, yet they retain the exceptional high-temperature characteristics that render HEAs appealing for aerospace applications \cite{nataraj2021systematic}. Understanding their phase stability and overall thermodynamic properties is therefore of broad interest by the community.

High Entropy Diborides (HEDBs) represent a significant subclass of Ultra-High Temperature Ceramics (UHTCs), engineered to withstand extreme operational conditions. Characterized by their ultra-high melting points, these materials exhibit exceptional thermal and electrical conductivity, alongside superior hardness, and outstanding resistance to wear and oxidation. A focal point of current research within this domain is the (Cr,Hf,Mo,Ta,Ti,Zr)B$_2$ diboride system. The application of the high entropy concept to this system facilitates the stabilization of a single Hexagonal Close-Packed (HCP) phase, mirroring the behavior observed in its constituent elements. Such stabilization not only augments hardness but also significantly enhances oxidation resistance, as documented in recent studies \cite{gild2016high,zhang2021ultra,mayrhofer2018high,feng2021superhard}. The strategic exploration and manipulation of high entropy compositions within the (Cr,Hf,Mo,Ta,Ti,Zr)B$_2$ framework open up new avenues for customizing and advancing the properties of UHTCs, offering groundbreaking solutions for applications requiring robust high-temperature performance.

\section{Methods}
\label{sec:methods}

\subsection{DFT data generation}

The training dataset generation involves acquiring a series of distinct structures, utilizing the superstructure enumeration methodology as integrated within the ICET framework\cite{aangqvist2019icet}. Given the exponential increase of feasible structures with the enlargement of the supercell size $n$ (the factor by which the supercell is larger than the primitive cell), a progressively larger random subset of these structures is selected for analysis. Specifically, Density Functional Theory (DFT) calculations are systematically performed for primitive (end-member) structures with $n=1$, and for binary combinations also with $n=2$. For larger configurations, where $n>3$, only a representative subset is subjected to DFT analysis. These structures undergo full relaxation via the Vienna Ab initio Simulation Package (VASP), and the data from all successfully converged ionic steps constitute our DFT ground truth dataset.

To maintain computational tractability, the dataset is constrained to fewer than 500 structures, aligning with the typical scale of structures analyzed in reported multi-component Cluster Expansion (CE) models \cite{nataraj2021systematic,fernandez2017short,nataraj2021temperature}. The training of the model incorporates stringent accuracy criteria for each ionic step performed with the Vienna Ab initio Simulation Package (VASP), ensuring the precision of calculated energies, stresses, and forces. Specifically, the model employs 4,000 Kpoints per atom to enhance the basis set's resolution, sets an energy convergence threshold of $10^{-6}$ eV for self-consistent field (SCF) calculations, and adopts a convergence criterion of $10^{-5}$ eV for ionic relaxation processes.

\subsection{Data partition and GNN model training}

In a manner consistent with the aforementioned approach, all data points for supercells with $n=1,2$, along with their respective ionic steps, are incorporated into the training dataset. For structures characterized by $n>3$ or greater, the dataset is divided into training, validation, and testing subsets. The selected model for this study is a Graph Neural Network, as developed by Chen et al. within the M3GNet framework \cite{chen2022universal}. The datasets for training, validation, and testing encompass all ionic step calculations from the same relaxation sequence. This strategy ensures the model's ability to evaluate complete relaxation results without having encountered any identical configuration structures during its training phase.

After iteratively trained on the combined training and validation datasets, the model then undergoes a final evaluation using the testing dataset. This last assessment involves a comprehensive relaxation process facilitated by the FIRE algorithm \cite{bitzek2006structural}, as integrated within the MatGL framework. This step is crucial for verifying the model's predictive accuracy and its ability to generalize across unseen data by simulating realistic structural relaxations.

\subsection{Cluster Expansion training and Monte-Carlo simulations}

Once the M3GNet GNN model for a given system of interest was obtained, we proceeded to sample the Cluster Expansion (CE) model training dataset. The sampling is constrained by the combinatorial expansion of unique structures within the crystal system, necessitating a judicious approach to computation. For structures with size $n>5$, M3GNet is employed to analyze only a subset of each set of structures.This strategy has the benefit of leveraging the constraints imposed by Density Functional Theory (DFT) calculations while enabling the examination of mixing energies across a substantially broader array of structures---extending into the tens of thousands! 
Despite the enhanced speed from the surrogate model, the configurational space remains too vast to sample exhaustively. Nevertheless, this approach facilitates the relaxation of an extensive number of structures—7,000 High Entropy Diborides (HEDBs) and more than 10,000 Body-Centered Cubic (BCC) High Entropy Alloys (HEAs)—demonstrating the model's efficiency in navigating and examining significant portions of the material design space.

The relaxed structures underwent a screening process to identify and exclude any instances of over-relaxation yielding heavily distorted structures incompatible with the original lattice with the standards put forth by Nataraj et al. \cite{nataraj2021systematic}. 

\begin{equation}
\small
\begin{aligned}
C=\left(\frac{B}{\sqrt[3]{|B|}}\right)^{-1} \cdot\left(\frac{A}{\sqrt[3]{|A|}}\right) \\
D=\left\|\frac{C+C^{\top}}{2}-I\right\|
\end{aligned}
\end{equation}

Where $B$ and $A$ are the unrelaxed and relaxed lattice matrices respectively. The $C$ matrix defines the resulting differential between the two. Finally, the $D$ is the resulting Distortion matrix, where values above 0.1 are removed from consideration.

For the remaining structures, a consistent data partitioning strategy was applied: unary and binary systems were always designated as training structures, while the remainder of the dataset was subjected to a four-fold cross-validation (CV) analysis. Ultimately, a final Cluster Expansion (CE) model is derived by training on the complete dataset, ensuring comprehensive coverage and robustness of the model across the entire spectrum of analyzed structures.

Upon the calibration of the Cluster Expansion (CE) model, this model is subsequently utilized as a computational kernel within a series of Monte Carlo (MC) Canonical Ensemble simulations. Starting from a fixed composition, the MC steps iteratively swap atom positions to thoroughly explore the configurational space. With every iteration, the Cluster Expansion engine calculates the configurational energy and chooses to accept or reject the change based on the Metropolis criterion at a given temperature. After sufficient MC steps, the equilibrium configuration for each temperature was determined.

\section{Results and Discussion}
\label{sec:results}
\subsection{Structure generation}

Unique structures within the specified crystal system were systematically generated using the exhaustive enumeration method provided by the ICET package, which is based on the Hart and Forcade algorithm \cite{hart2008algorithm}. Subsequently, these structures were categorized by their size ($n$, representing the multiplier of the primitive cell size), and then selectively included in the DFT and M3GNet datasets. This selection process involved either incorporating all structures of a given size, $n$, or employing a random sub-selection for analysis. A notable advantage of using M3GNet for the relaxation of structures is its capacity to analyze a broader subset of the generated structures efficiently---due to its much lower computational cost compared to DFT calculations. This capability is demonstrated in Tables \ref{tab:bcc_space} and \ref{tab:hcp_space}, where a larger number of structures can be processed compared to traditional methods based on DFT.

\begin{table}[!htbp]
\centering
\caption{BCC crystal structure generated structures and selected for the different relaxation engines.}
\label{tab:bcc_space}
\small 
\begin{tabularx}{\columnwidth}
{@{}lXXXX@{}} 
\toprule
\makecell{Size of primitive\\ units} & Size & No. of unique str. & DFT & M3GNet \\
\midrule
1 (endmembers) & 1 & 6 & 6 & 6 \\
2 & 2 & 30 & 30 & 30 \\
3 & 3 & 150 & 75 & 150 \\
4 & 4 & 1350 & 75 & 1350 \\
5 & 5 & 3984 & 75 & 3984 \\
6 & 6 & 42200 & 100 & 4000 \\
7 & 7 & 130800 & 125 & 4314 \\
\bottomrule
\end{tabularx}
\end{table}

\begin{table}[!htbp]
\centering
\caption{HEDB crystal structure generated structures and selected for the different relaxation engines.}
\label{tab:hcp_space} 
\small 
\begin{tabularx}{\columnwidth}{@{}lXXXX@{}} 
\toprule
\makecell{Size of primitive\\ units} & Size & No. unique str. & DFT & M3Gnet \\
\midrule
1 (endmembers) & 3 & 6 & 6 & 6 \\
2 & 6 & 45 & 30 & 45 \\
3 & 9 & 250 & 45 & 250 \\
4 & 12 & 2280 & 60 & 1500 \\
5 & 15 & 6174 & 60 & 2000 \\
6 & 18 & 80180 & 60 & 2000 \\
\bottomrule
\end{tabularx}
\end{table}

The structures designated for relaxation via VASP were optimized by slightly modifying the atomic positions and lattice vectors at every electronic relaxation step, which in turn consists of a self-consistent iterative process that if successful returns an electronic ground state. Once the energy is obtained, the forces and stresses are also calculated to inform the next step of the ionic relaxation. Accordingly, energy, forces, and stresses from those self-consistent iterations achieving an energy convergence threshold of $10^{-6}$ eV within fewer than one hundred steps are recorded.

\subsection{M3GNet model training}

The dataset, as generated in the preceding section, is composed of converged self-consistent calculations that include a structure as the input, and energy, forces, and stresses as the outputs. Within this collection, structures may either originate from the same ionic relaxation process (identical initial state) or from disparate ones (originating from different enumeration methods as their initial state)---this distinction is important in the strategy for partitioning data for model training and validation. Consequently, data partition for model training and validation is done by taking this into account, and we group structures that belong to the same ionic relaxation path by partitioning the train/validation/test datasets by the initial uniquely enumerated structure they were generated from.

The M3GNet architecture is implemented as in the MatGL package \cite{chen2019graph, chen2021learning, chen2022universal}. The training consists of 100 iterations for which the energy, forces, and stresses are fitted to the M3GNet hyperparameters. This GNN-based model consists of an initial graph converter that computes the three-body interactions in combination with the geometrical and atomic profiles of the constituent elements. To corroborate the model's accuracy in the crystal systems of interest, we perform a final test comparison. In this analysis, we juxtapose the test dataset outcomes---specifically, comparing results derived from Density Functional Theory (DFT) against those predicted by the pre-trained model and another model that has been retrained with the crystalline and atomic systems of interest.

\begin{figure*}
    \centering
    \includegraphics[width=\textwidth]{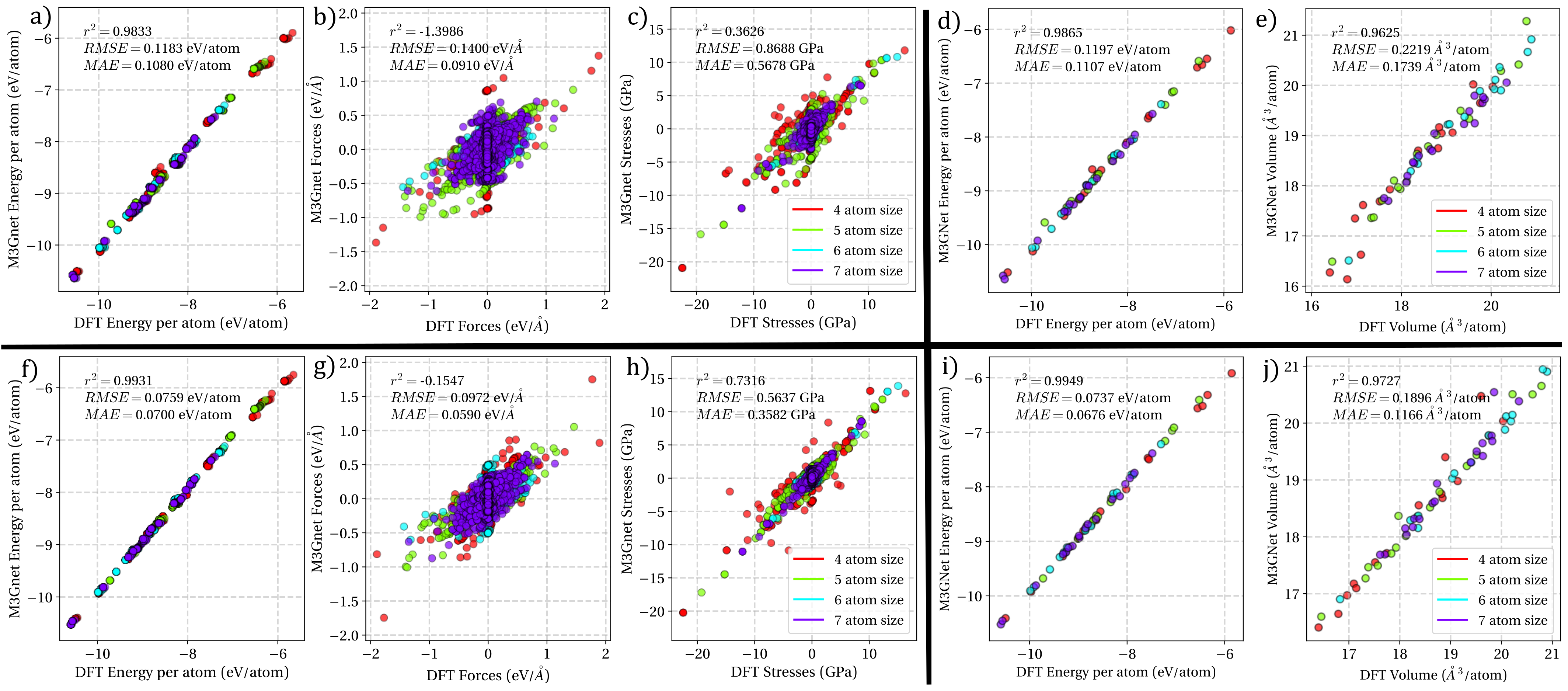}
    \caption{\textbf{Self-consistent and relaxation parity comparison for the BCC system.} Pre-trained (top row) and system-specific trained (bottom row) models comparison. In a), b), and c) we compare the energies, forces, and stresses respectively, of self-consistent calculations for ionic steps using the M3GNet available pretrained model. Similarly, for f), g), and h) for the newly in-house trained model. The ionic relaxation task accuracy is evaluated by comparing the last ionic step in both the M3GNet model and in VASP when starting from the same initial structure: in d), and e) for the final energy and volume for the pre-trained model. And in i) and j) for the in-house M3GNet model trained in only AlHfNbTaTiZr BCC structures.}
    \label{fig:bcc_par}
\end{figure*}

Figure \ref{fig:bcc_par} illustrates the validation of the model for the BCC system using the test dataset, encompassing both self-consistent calculations (ionic steps) and ionic relaxations. The initial three columns display the self-consistent results, including energy per atom, forces, and stresses. Notably, the predicted energy for the electronic ground state aligns closely with results obtained from VASP, indicating accurate model performance. However, further retraining on the crystalline BCC system does not significantly enhance this accuracy. There is a slight improvement in the predictions of forces and stresses, which is beneficial for the ionic relaxation process as it relies on accurately determining the forces acting on each atom to achieve a geometrically relaxed state. Despite this, the model's force prediction accuracy remains poor post-retraining, but it's important to note that the Root Mean Square Error (RMSE) for these predictions has substantially decreased, and the stress prediction accuracy has improved significantly. The good-to-excellent accuracy in the energy and stress predictions suggests that there may be beneficial cancellation of errors in the predicted forces.

\begin{figure*}
    \centering
    \includegraphics[width=\textwidth]{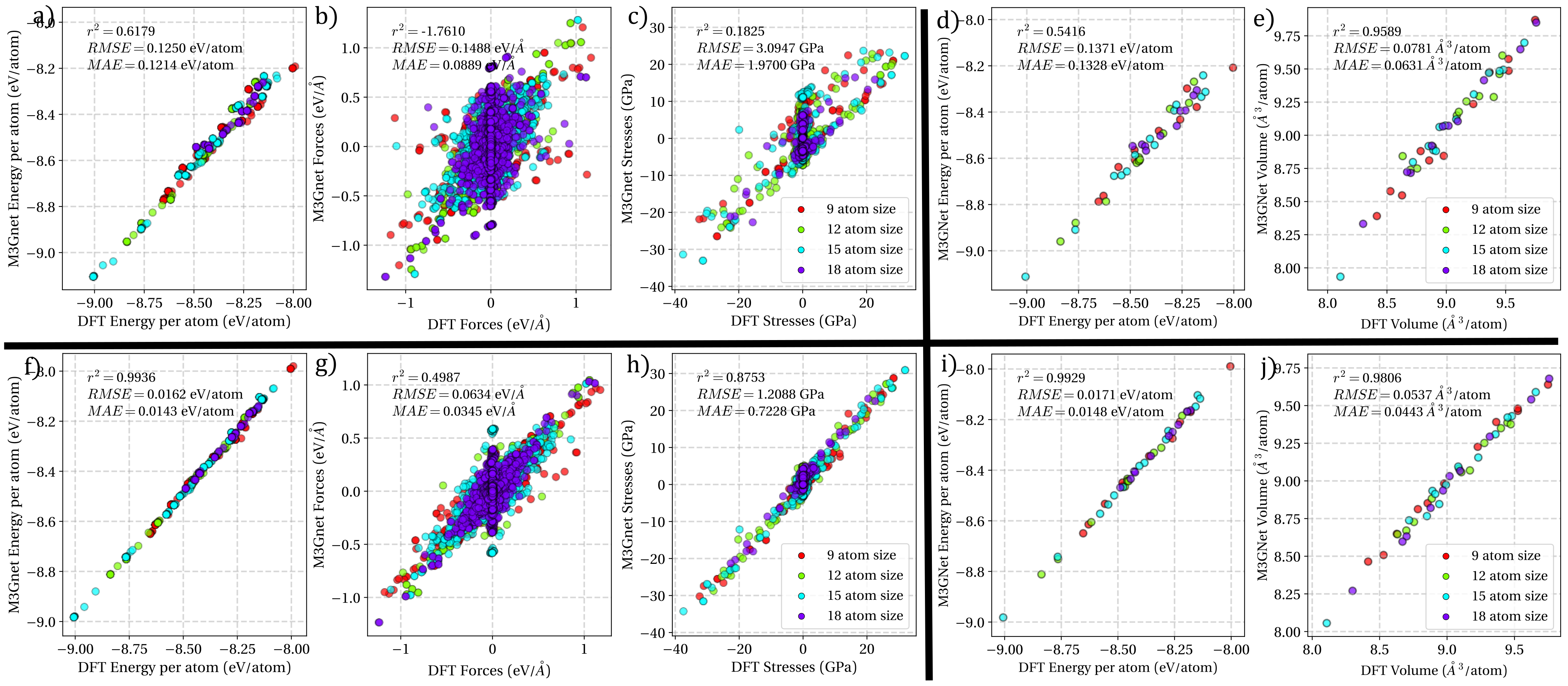}
    \caption{\textbf{Self-consistent and relaxation parity comparison for the HEDB system.} Pre-trained (top row) and system-specific trained (bottom row) models comparison. In a), b), and c) we compare the energies, forces, and stresses respectively, of self-consistent calculations for ionic steps using the M3GNet available pretrained model. Similarly, for f), g), and h) for the newly in-house trained model. The ionic relaxation task accuracy is evaluated by comparing the last ionic step in both the M3GNet model and in VASP when starting from the same initial structure: in d), and e) for the final energy and volume for the pre-trained model. And in i) and j) for the in-house M3GNet model trained in only (CrHfMoTaTiZr)B$_2$ HEDB structures.}
    \label{fig:hcp_par}
\end{figure*}

Figure \ref{fig:hcp_par} presents the results for the High Entropy Diboride (HEDB) system, where retraining the M3GNet model specifically on this crystalline system significantly enhances its predictive accuracy. The improvements are evident in the self-consistent predictions of energy, forces, and stresses, and the accuracy of ionic relaxation also sees considerable gains, with a Root Mean Square Error (RMSE) around $0.1~eV/atom$ for the fully relaxed states. This improvement is likely due to the larger crystalline system size involved: unlike the pre-trained model, which mainly learned from smaller, ordered compounds, this study incorporates larger structures from specific enumerated states in the HEDB system. Thus, retraining the M3GNet model is essential for its successful application as a surrogate relaxer for this particular system.

The final columns in Figures \ref{fig:bcc_par} and \ref{fig:hcp_par} display the system's final volume after relaxation. These columns highlight the model's capability to accurately predict the volume at the final relaxed ionic step, crucial for correctly extrapolating the energies specific to the crystal system of interest for the Cluster Expansion (CE) method. The observed discrepancies in ionic relaxation results can be attributed to the utilization of different optimization methods: the Conjugate Gradient method in the VASP package versus the Fast Inertial Relaxation Algorithm (FIRE) implemented in the ASE package \cite{larsen2017atomic}. Notably, for both the BCC and HEDB systems, the in-house trained M3GNet model demonstrates high accuracy, with the $R^2$ parameter for predicting the energy and volume of the relaxed ionic state exceeding 0.96.

\subsection{M3GNet model High-Throughput relaxation} 

After validating the model with unseen structures, we expanded the application of our surrogate relaxer to encompass a broader array of structures. This approach mirrors the procedure used in Density Functional Theory (DFT) calculations, but with the significant advantage of much faster relaxation times, thus mitigating the limitations imposed by computational costs associated with DFT calculations. However, the selection process remains influenced by the combinatorial complexity inherent in the number of enumerated configurations in any lattice system with a large number of constituents.

From the comprehensive structure generation detailed in Tables \ref{tab:bcc_space} and \ref{tab:hcp_space}, a subset is chosen for relaxation via the M3GNet model, following a methodology akin to that employed in previous DFT analyses. This involves either encompassing the entire subspace of a given size \(n\) or randomly selecting a proportion of configurations at that size. Despite the reduction in relaxation costs, the sheer volume of possible configurations exceeds our computational capabilities. Consequently, while we still limit our selection to certain enumerated structures for larger sizes, the efficiency gains allow us to consider a more extensive selection than previously feasible.

\begin{figure*}[hbt]
    \centering
\includegraphics[width=0.8\textwidth]{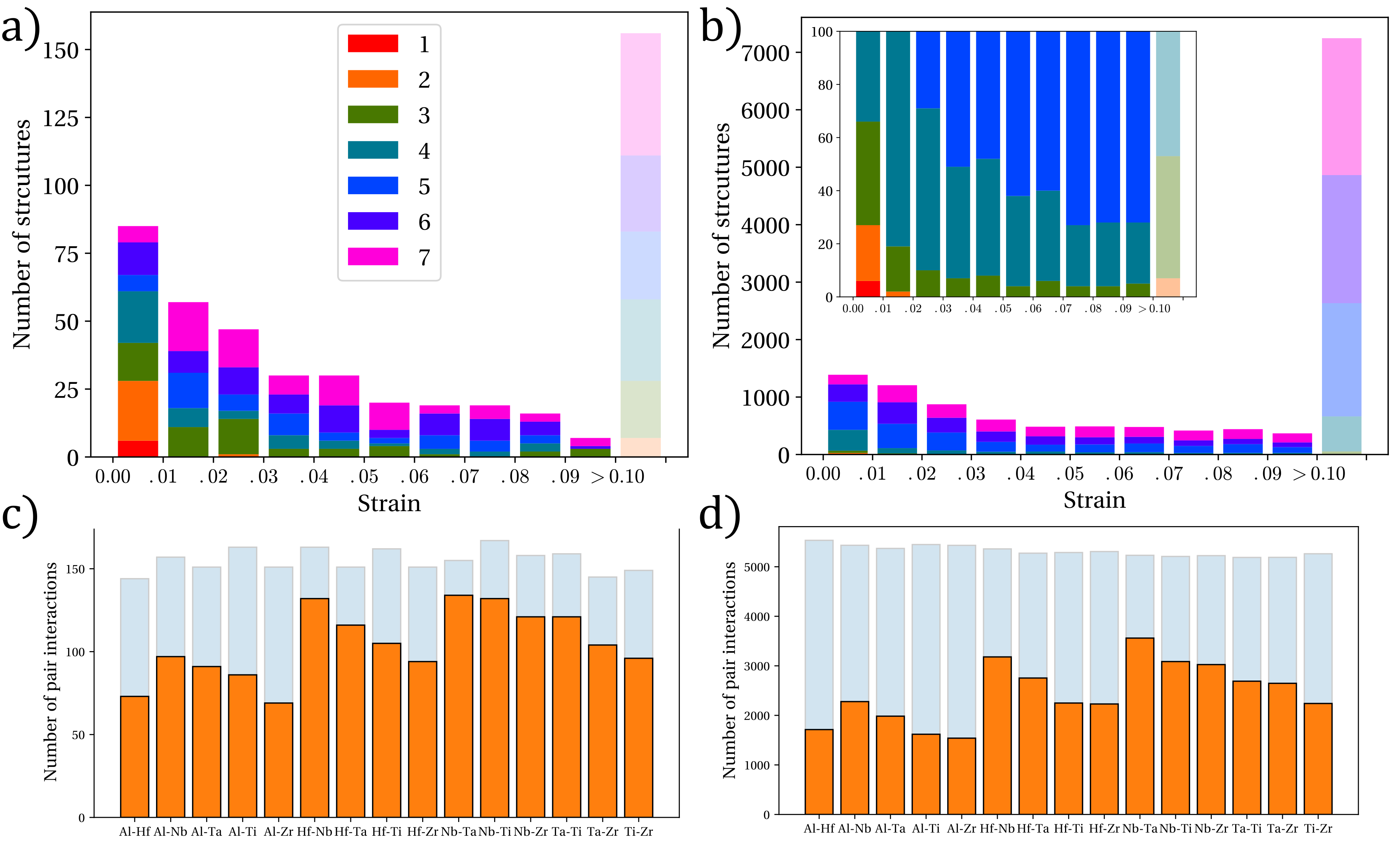}
    \caption{\textbf{BCC System Structure Analysis Overview.} This figure presents a dual-level analysis focusing on strain distribution and pair interactions within the BCC system. Top row: Visualization of strain distribution across varying structure sizes, with color indicating structure count and x-axis representing strain intensity in the relaxed state, comparing (a) DFT and (b) M3GNet results. Bottom row: Analysis of pair interactions in the dataset, distinguishing between structures under a specific strain threshold (solid areas) and the entirety of the dataset (shaded areas), showcased through (c) DFT and (d) M3GNet findings. This comprehensive analysis elucidates the comparative accuracy of DFT versus M3GNet models in simulating strain and interaction dynamics.}
    \label{fig:structs_bcc}
\end{figure*}

The relaxation was carried out using M3GNet as the electronic structure predictor engine and the FIRE algorithm as the ionic relaxer. We ran all structures in the designated sub-selection with its respective model. In Figure \ref{fig:structs_bcc} the results for both DFT and M3GNet relaxation are showcased, we also notice the difference in numbers at the two followed approaches. For both, we analyze the relaxed state and calculate the strain on them. As discussed earlier, if the strain (distorsion) is greater than the set threshold of 0.10, the structure is no longer considered BCC which implies a loss of information for the CE method (shown as the shaded area in Figure \ref{fig:structs_bcc}). The regressor model approach circumvents this issue by extrapolating both information from ionic steps within and outside the crystalline system. Therefore, the final HT exploration of the system using M3GNet has to follow this same constraint, but the scale makes the M3GNet approach contain much more information even if there are still over-relaxed structures.

\begin{figure*}[hbt]
    \centering
    \includegraphics[width=0.8\textwidth]{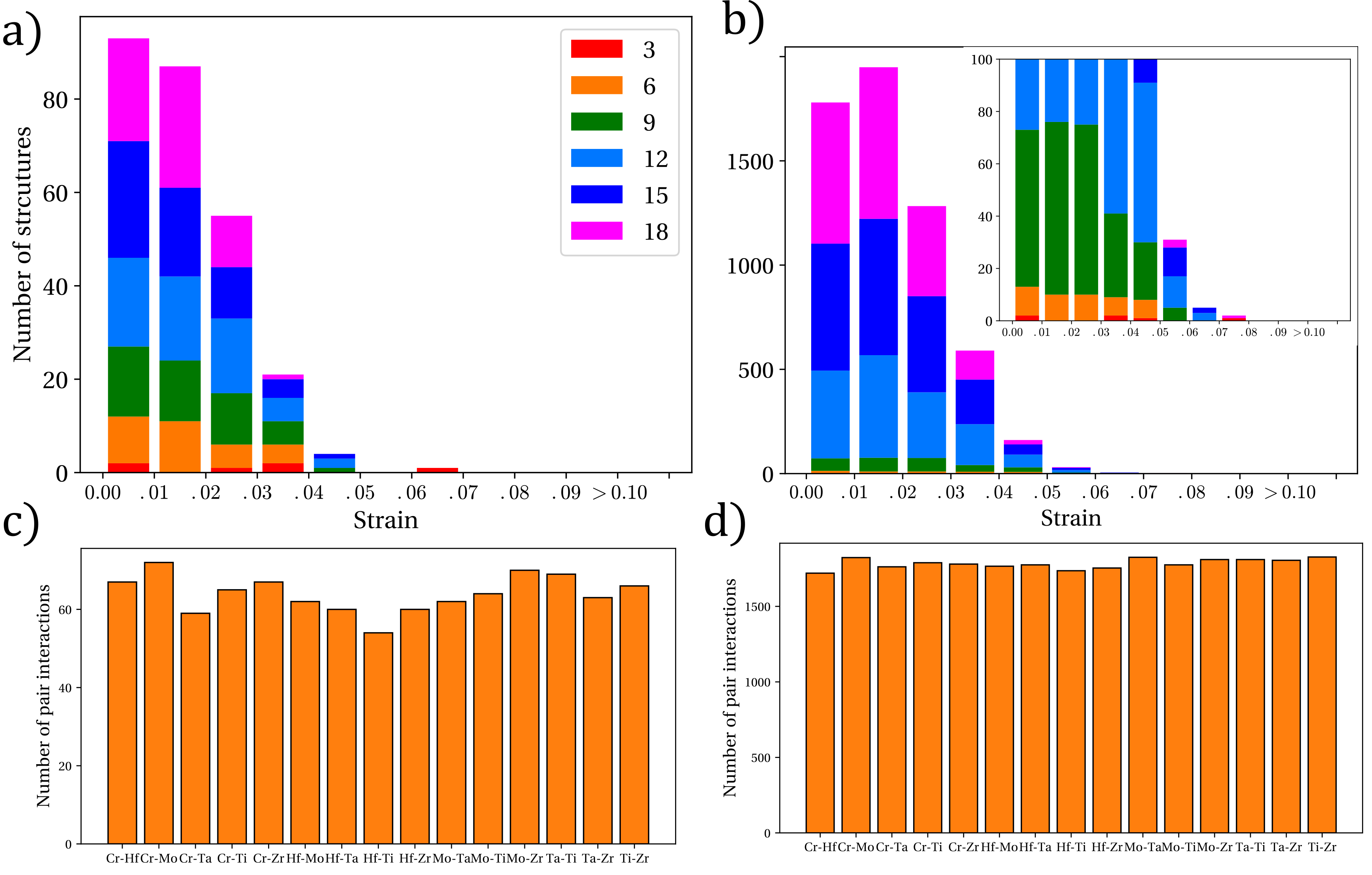}
    \caption{\textbf{Detailed Structure Analysis of the HEDB System.} The figure offers a nuanced look at strain distribution and pair interactions within the HEDB system. The top row illustrates strain distribution across varying structure sizes, with structure count denoted by color on the y-axis and strain intensity in the relaxed state on the x-axis. The atomic size is scaled in multiples of three, equivalent to the size of the diboride primitive unit cell, comparing (a) DFT and (b) M3GNet results. The bottom row quantifies the number of pair interactions, presenting (c) DFT and (d) M3GNet outcomes. This juxtaposition highlights the precision of DFT versus M3GNet in capturing structural dynamics and interaction density within the dataset.}
    \label{fig:structs_hcp}
\end{figure*}

In the case of the High Entropy Diboride (HEDB) system, Figure \ref{fig:structs_hcp} illustrates a similar analysis, revealing that the diboride system exhibits greater mechanical stability. Notably, no structures exceed the strain threshold post-relaxation with both DFT and M3GNet, with the latter unsurprisingly mirroring this pattern of stability. 

For both the HEDB and BCC systems, structure size does not present a definitive trend, yet a decrease in the frequency of structures is observed as strain levels escalate. Specifically, within the BCC system, strains accumulate beyond the threshold, leading to a predominance of strained structures over those observed in the high-throughput M3GNet relaxation process.

\begin{figure*}
    \centering
\includegraphics[width=0.95\textwidth]{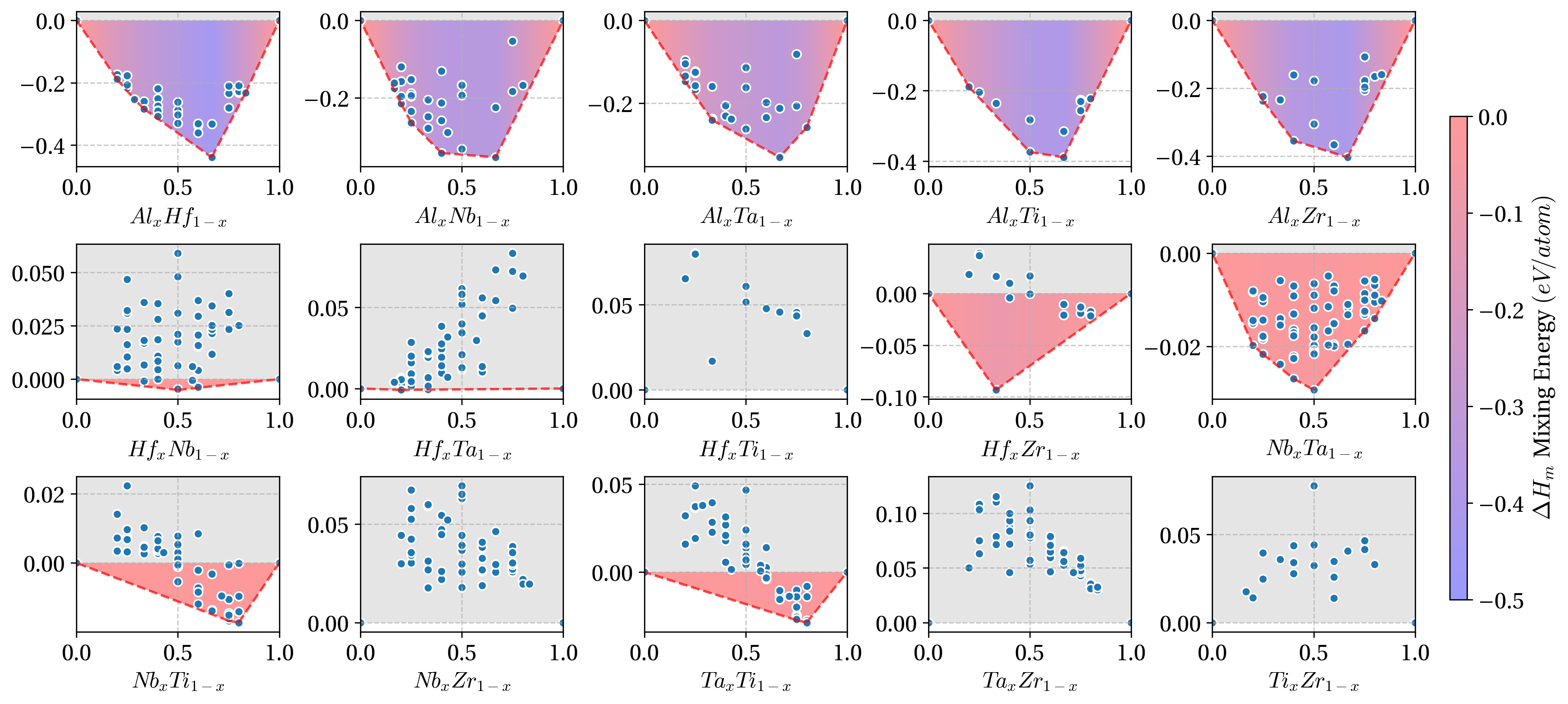}
    \caption{\textbf{Convex-Hull Analysis for the BCC System.} Formation enthalpies at 0K as predicted by the trained M3GNet model are shown for every binary in the system. A red enclosed area, indicating compositions where the Convex-Hull deviates from a zero line, highlights energetically favorable formations. The color intensity within this area reflects the degree of negativity in formation energy, providing insight into the stability of various compositions.}
    \label{fig:hull_bcc}
\end{figure*}

\begin{figure*}
    \centering
\includegraphics[width=0.95\textwidth]{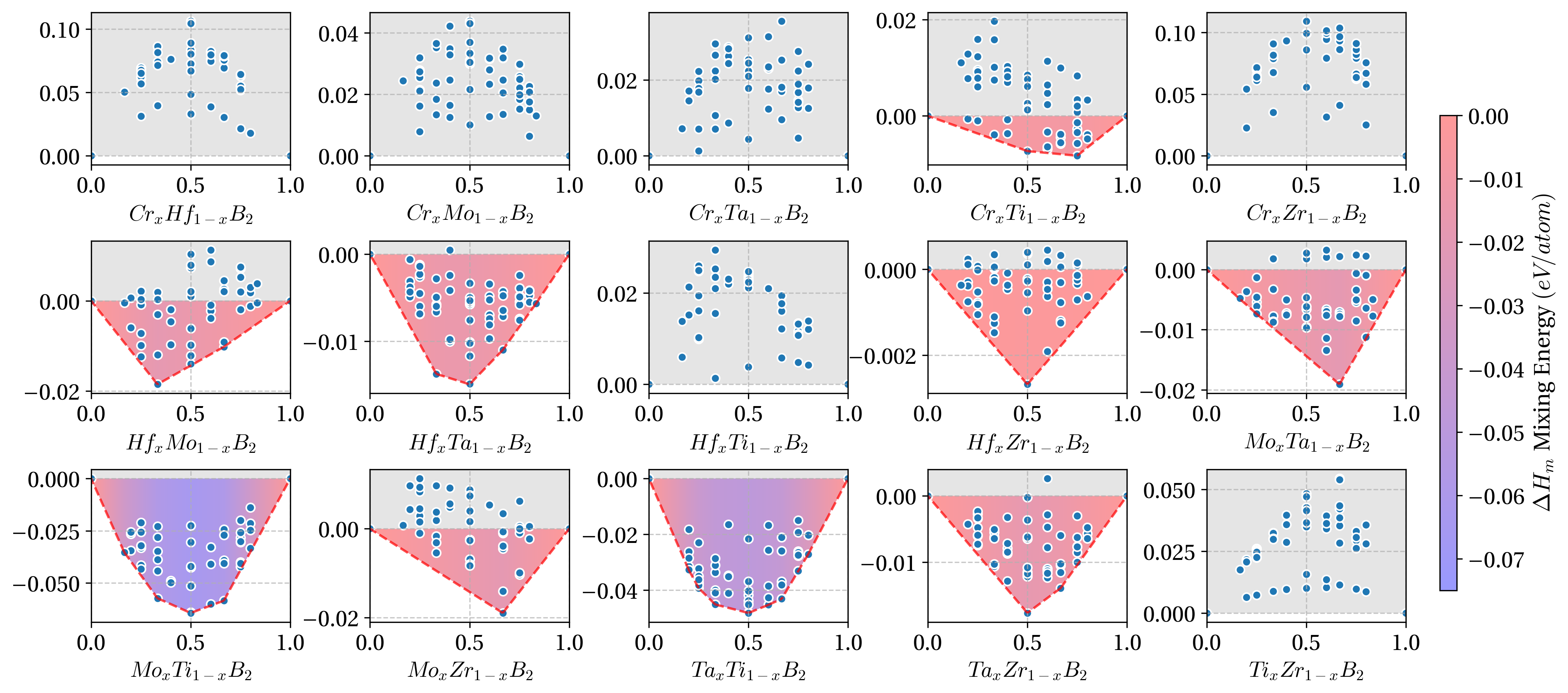}
    \caption{\textbf{Convex-Hull Analysis for the HEDB System.} Formation enthalpies at 0K, as predicted by the trained M3GNet model, are presented for every pseudo-binary in the system. A distinct red enclosed area indicates compositions where the Convex-Hull deviates from a zero line, signifying energetically favorable formations. The intensity of the red coloring correlates with the degree of formation energy negativity, illustrating the stability landscape of various compositions within the system.}
    \label{fig:hull_hcp}
\end{figure*}

The Convex-Hull analysis, requiring a large number of calculations, becomes significantly more feasible with the aid of the trained regressor. This analysis may target exclusively structures within a specific crystalline system or extend to any structure comprising the relevant elements---in the latter case, the construction would correspond to a better approximation to the true ground state of the system. Adhering to established protocols, we focus solely on structures relaxed within their native crystalline systems. This approach aims to ascertain whether a particular binary system exhibits a stable energy landscape relative to its constituent unmixed states, rather than identifying the true ground state across all possible crystal systems.

Initially, we examine the Body-Centered Cubic (BCC) High Entropy Alloy (HEA) system results. Due to over-relaxation leading to information loss, not all generated structures are included in our analysis; we consider only those depicted with solid coloring in Figure \ref{fig:structs_bcc}a. For these selected structures, we calculate the mixing enthalpy, $\Delta H_m$, by comparing their total energies—derived from the M3GNet-relaxed unaries—to maintain energy consistency. The mixing enthalpies (in units of eV per atom) for each binary system are shown in Figure \ref{fig:hull_bcc}. Notably, binaries containing aluminum and the Nb-Ta pair consistently exhibit negative $\Delta H_m$ values across all compositions, suggesting their solid solutions may prefer mixing rather than phase separation.  Furthermore, the most energetically favorable combinations, such as Al-Zr or Nb-Ta, may tend to cluster. Nb-Ti and Nb-Ta show comparable mixing enthalpies, suggesting stable solid solutions.

The analysis was extended to the High Entropy Diboride (HEDB) system, with a notable distinction that pair interactions incorporate a complete formula unit for diborides. Thus, each binary combination in this system is accompanied by a corresponding quantity of boron atoms. The high-throughput (HT) relaxation of various structures accounts for both direct and screened interactions, whether within the same plane or across different planes, enriching the study of binary compositions. In Figure \ref{fig:hull_hcp}, we illustrate the boride pseudo-binaries. Notably, the most exothermic pseudo-binary systems correspond to Mo-Ti, Ta-Ti, although most systems have relatively low enthalpies of mixing (of either sign). Our analysis suggests that Cr does not generally contribute to stable pseudo-binaries, with the exception of Cr-Ti, which is noteworthy since Ti is also a component of energetically favorable binaries like Mo-Ti, and Ta-Ti. It suggests that the preference for specific atom clustering is influenced not solely by the most stable binary interactions but also by other competing binary and more complex interactions. Nevertheless, binary subsystems serve as valuable indicators of phase stability.

\subsubsection*{M3GNet model time efficiency}

\begin{table*}[!htbp]
\centering
\footnotesize 
\caption{Time comparison for DFT vs M3GNet-FIRE relaxation for HEDB system.}
\label{tab:time_hcp}
\begin{tabular}{cccccc}
\toprule
Size & \# of Struct. & DFT Tot. (s) & DFT Avg. (s) & M3GNet Tot. (s) & M3GNet Avg. (s) \\
\midrule
3    & 6   & 606.76 & 101.13 & 4.23  & 0.70 \\
6    & 30  & 5038.25& 167.94 & 48.25 & 1.61 \\
9    & 45  & 11440.98& 254.24& 94.85 & 2.11 \\
12   & 60  & 18657.2 & 310.95& 171.43& 2.85 \\
15   & 60  & 32529.53& 542.49& 327.54& 5.46 \\
18   & 60  & 35018.96& 583.65& 421.80& 7.03 \\
\midrule
Total& 261 & 103311.67& 395.83& 1068.10& 4.09 \\
\bottomrule
\end{tabular}
\end{table*}

\subsection*{Cluster Expansion}

As mentioned in the Introduction, the cluster expansion is a method in the statistical mechanics of lattice systems that defines a quantity of interest as a linear regression of interacting terms that are subsequently truncated at a reasonable level of accuracy for the application. The exact general form is shown as equation \ref{eqn:general_CE}.
\begin{equation}\label{eqn:general_CE}
H(\sigma)=\sum J_f \Pi(\sigma)_f
\end{equation}

Where $H(\sigma)$ would be the true value of our measured quantity as a function of descriptors $\sigma$. However, in our specific case of determining the energetics of a crystal system, the equation expands to form equation \ref{eqn:energy_CE}. Where $J$ is a fitted coefficient vector or Effective Cluster Interaction (ECI).

\begin{equation}\label{eqn:energy_CE}
\mathrm{E}(\sigma)=J_0+\sum_i^{n_{\text {sites }}} J_i \sigma_i+\sum_{i, j}^{n_{\text {pairs }}} J_{i, j} \sigma_i \sigma_j+\sum_{i, j, k}^{n_{\text {triple }}} J_{i, j, k} \sigma_i \sigma_j \sigma_k+\cdots
\end{equation}

Specifically, each summation term, including the $J_o$ term, physically denotes the energies of separate cluster sets. Typically these would be sourced from DFT, but recall that for computational efficiency we are using results produced by M3GNet to sample a  well-populated and dispersed feature space.

With the assumption that the ECIs are properly fitted (trained) to the desired phase space, following the standards originally put forth by Axel van de Walle et al. \cite{vandewalle2004automating} and using the Least Absolute Shrinkage and Selection Operator (LASSO) as implemented in the ICET package\cite{aangqvist2019icet}, we have a model capable of calculating the energies of arbitrary atomic configurations on a lattice at a precision close to that of DFT calculations, but at the much lower (by orders of magnitude) cost of running the re-trained M3GNet model. The cluster expansion model can then be used to extrapolate the energy of any structural configuration and be used to investigate the phase stability and the thermodynamics of short-range ordering (SRO) in compositionally complex systems.

The training space for the BCC system consists of all clusters up to $3 \cdot a_0$, $2 \cdot a_0$, and $1.5 \cdot a_0$ in pair, triplets, and quadruplets, respectively. In addition, empty and point clusters are also added to the training feature space. Therefore, the total initial feature space contains 1, 1, 15, 17, and 7 empty, point, pair, triplet, and quadruplet figures each expanding to 1, 5, 225, 1485, and 1260 clusters. Finally, the total number of features in the input vector contains is 2,976. 

\begin{figure}[!htbp]
    \centering
\includegraphics[width=\columnwidth]{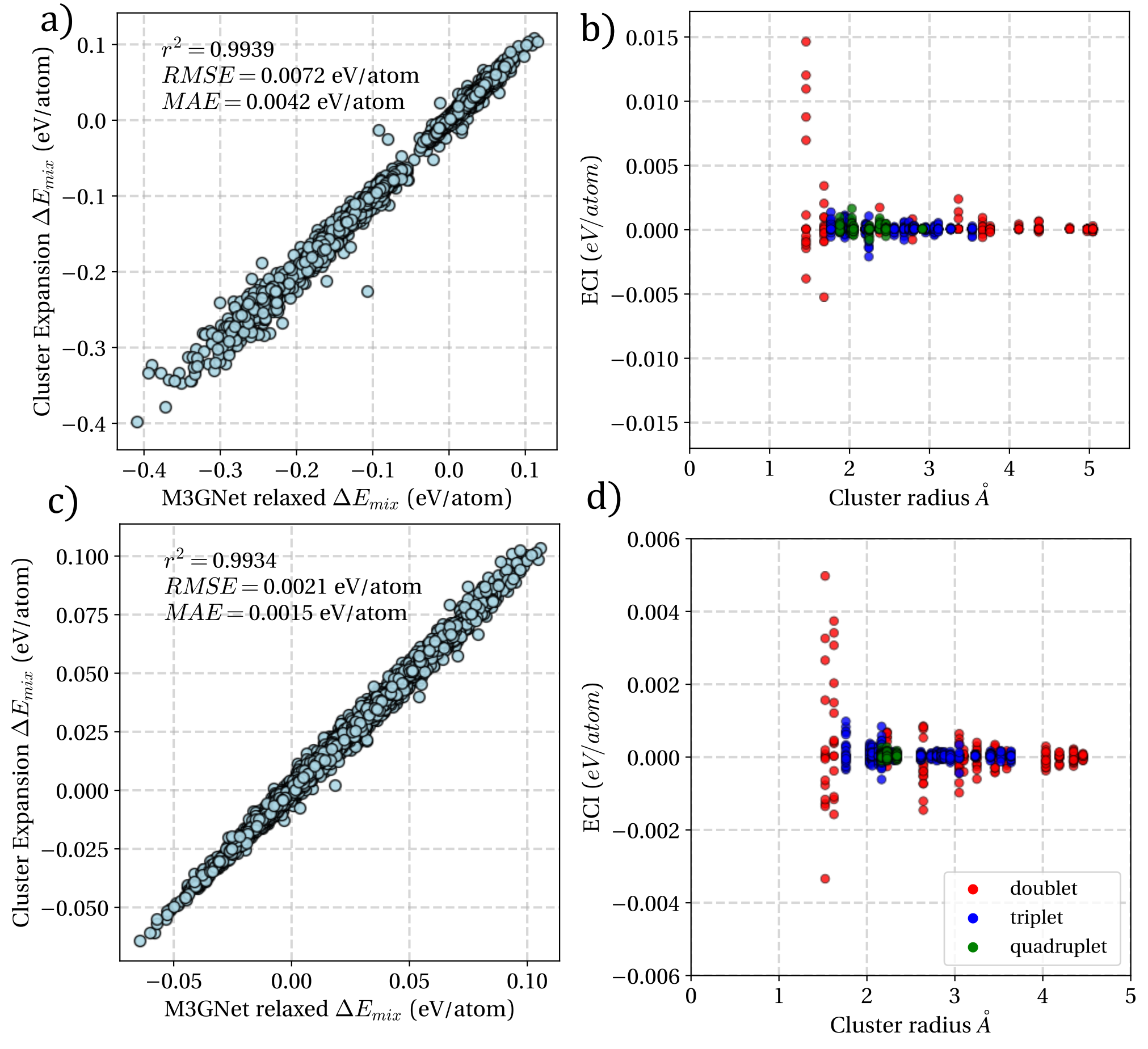}
    \caption{\textbf{4-fold CV Results for Cluster Expansion vs. Relaxed GNN-FIRE Structures}. (a) BCC system results, showing CE performance within tolerable RMSE values, well below the 10 meV/atom threshold for a successful model. (b) ECI values vs. cluster radius for BCC. (c) HEDB system results, with similar tolerance levels. (d) ECI values vs. cluster radius for HEDB, indicating CE's effectiveness across different systems.}
    \label{fig:ce}
\end{figure}

As shown in Figure \ref{fig:structs_bcc}a, the total number of structures still considered BCC is filtered down to 6,754. From these, all structures up to atomic size 3 are allocated into the training data and the rest is divided into 4 equally split random subsets. Consequently, 4 different datasets for each system are created, each containing almost 25\% of the data. Cross-validation results for the BCC CE model are shown in Figure \ref{fig:ce}a, the 4-fold CV shows consistent results in all the relatively large validation sets. After this, the final CE model is obtained by training in all relaxed data still within the system: this model still contains 656 non-zero parameters by optimizing via the sparsifying LASSO method.

In turn, Figure \ref{fig:structs_bcc}b shows the CV the 4-fold method for the HEDB system. The figure shows that the cluster expansion is able to predict the energies of atomic configurations within less than 2 $meV/atom$. In this case, we include up to $3 \cdot a_0$, $2.2 \cdot a_0$, and $1.5 \cdot a_0$ in pairs, triplets, and quadruplets, respectively, where $a_0$ is the smaller lattice parameter. Which in length covers 14 pairs, 17 triplets, and 3 quadruplets. Then, the final cluster space is divided into 1, 5, 195, 1405, and 875 clusters of ascending order making up 2,481 features. Like before, after the validation of the method is proved for this system the definitive CE model is obtained by training in all data, which ends up with 1,392 non-zero parameters.

\subsection{SRO calculation via Canonical Monte Carlo Simulations}

Upon developing a M3GNet-trained Cluster Expansion (CE) model for both systems under investigation, we utilized it to examine the ordering tendencies at equiatomic concentrations across a temperature spectrum ranging from 0K to 3,200K, with intervals of 50K through Monte Carlo calculations. We used the Canonical Ensemble $(N_i,V,T)$to carry out the calculations at constant composition, volume, and temperature. During each iteration, a trial swap between two distinct atoms is attempted and the acceptance probability is estimated as a function of the energy change resulting during the swap---$P = min \{1, exp\left[-\Delta E/k_BT \right] \}$. At lower temperatures, the transition to a state of lower energy has a probability of one, suggesting that, given sufficient iterations, the system is likely to achieve the most energetically favorable configuration within this supercell.

The Short Range Order (SRO) parameters are calculated at every 50 steps, and the values belonging to the last 20,000 steps are averaged, leaving a burn-in period of 30,000 trials. We use the well-known Warren-Cowley SRO \cite{cowley1965short} parameters as expanded to a multi-component system by de Fontaine \cite{de1971number} which can be reduced to Eq. \ref{eqn:sro} \cite{rao2022analytical} where the SRO parameter, $\alpha_{ij}^{(r)}$ is only proportional to the probability of finding an $(i,j)$ bond at a distance $(r)$:

\begin{equation}
    \alpha_{ij}^{(r)} = 1- \frac{P_{ij}^{(r)}}{2c_ic_j}
    \label{eqn:sro}
\end{equation}

Within this formulation, negative Warren-Cowley SRO parameters ($\alpha_{ij}^{(r)}$) imply $P_{ij}^r>2c_ic_j$, thus indicating that the probability of finding an $i$ at a distance $r$ from a $j$ atom is greater than what one would estimate assuming random lattice occupancy, indicating a tendency for both atoms/species to co-segregate within the lattice. A positive $\alpha_{ij}^{(r)}$, on the other hand, suggests an anti-co-segregation tendency---species $i$ and $j$ would tend to be as far away as possible from each other.

\begin{figure}[!bt]
    \centering
    \includegraphics[width=0.85\columnwidth]{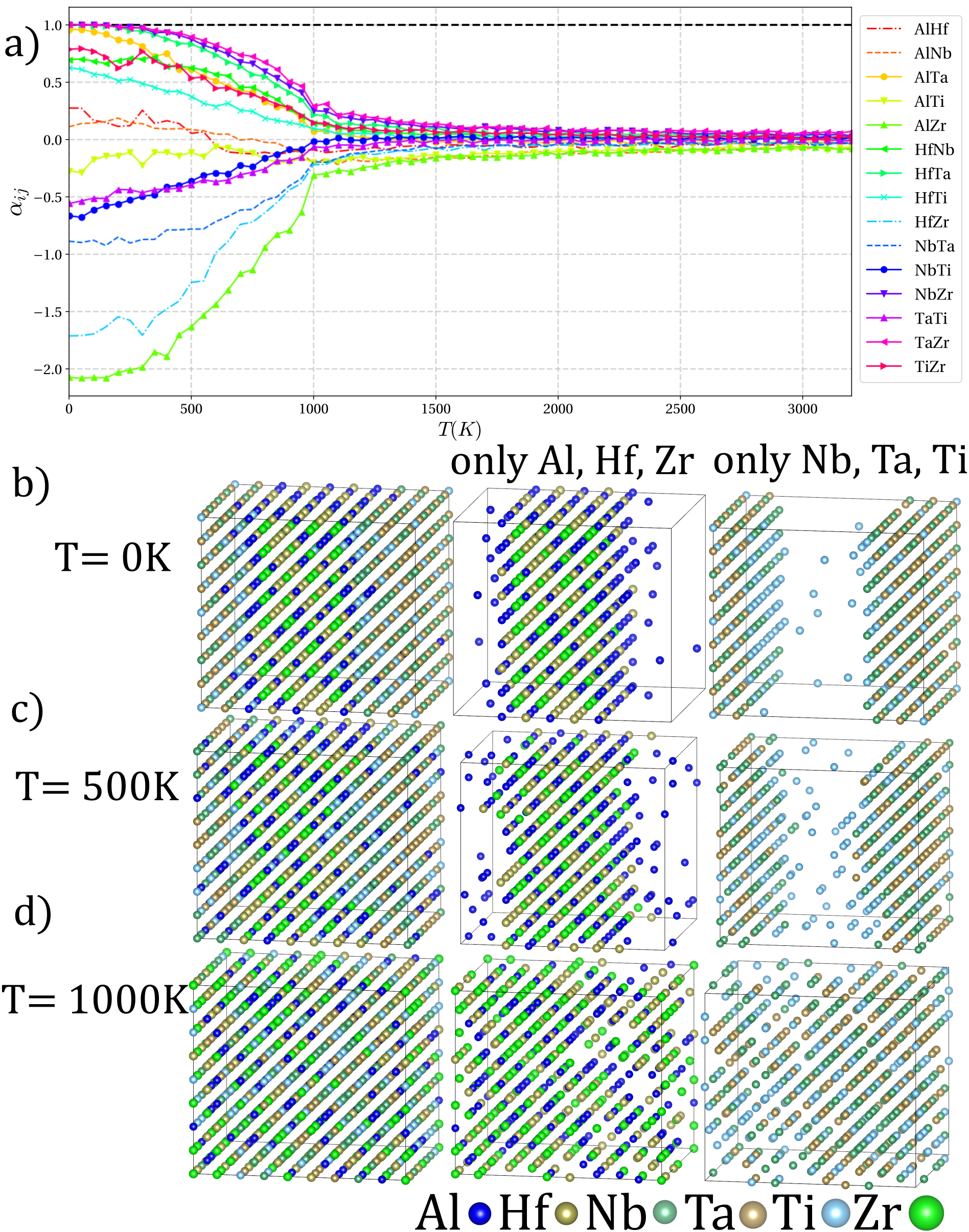}
    \caption{\textbf{SRO and Super-cell Graphs for the BCC System.} (a) Displays SRO curves for the first nearest neighbor interactions within the BCC lattice. Subfigures (b), (c), and (d) present graphical representations of the super-cell configuration at the last MC step for temperatures of 0, 500, and 1000 K, respectively. The first column shows all constituent atoms, while subsequent columns highlight atoms that cluster together.}
    \label{fig:sro_bcc_full}
\end{figure}

In the Body-Centered Cubic (BCC) system, as depicted in Figure \ref{fig:sro_bcc_full}-a, the Short-Range Order (SRO) for the first nearest neighbor is presented. At high temperatures, the SRO values for all binary combinations converge to zero, indicating an absence of clustering or dispersion. Conversely, at lower temperatures, atomic affinities lead to a distinct ordering pattern. Notably, as shown in Figure \ref{fig:sro_bcc_full}-a, the interactions between Hf-Zr suggest clustering (see Figure \ref{fig:sro_bcc_full}-b), and together with Al-Zr, they form a notable aggregation of these three elements. However, a slight repulsion between Al and Hf at low temperatures positions them further apart (indicating 2nd neighbor interactions), while they both remain in close proximity to Zr. As the temperature increases, this relationship gradually diminishes, with Hf increasingly occupying larger regions of the supercell. 

Figure \ref{fig:sro_bcc_full}-a indicates that while there are strong ordering tendencies at low temperatures, the equilibrium atomic configuration---within a Canonical ensemble---begin to converge at relatively low temperatures (1,000~K), which is somewhat surprising given the relatively high melting point of this alloy. Despite the relatively unexpected results, they are in agreement with the work by Nataraj \emph{et al.}\cite{nataraj2021temperature}. Using a DFT-trained cluster expansion formalism, they investigated the same  AlHfZrNbTaTi BCC RHEA system. While they did not explicitly computed the evolution of the SRO W-C parameters as a function of temperature, they calculated the configurational entropy of the system. Their calculations suggest that by ~1,000~K, the configurational entropy is already $\sim$90\% of the maximum configurational entropy, $S^{conf}_{max}$, expected for this equiatomic composition (-$k_B \ln{1/6}\sim 1.8 k_B$). At 1,500~K, the configurational entropy is already $\sim$94\% of $S^{conf}_{max}$, with convergence to maximum entropy occurring at $\sim$1,500~K. These observations are in line with our calculations. An important consequence of this analysis is that, in this system, SRO is unlikely to serve as a useful alloy 'design knob' as the alloy is likely to be fully disordered at any relevant operating temperature---we note that vibrational entropy contributions would only depress the temperature at which the alloy turns random-like\cite{nataraj2021temperature}.

We also note that the reported configuration by Nataraj \emph{et al.}\cite{nataraj2021temperature} for the low temperature case matches the trend obtained in this work. Especially the cluster of the Nb-Ta pair -- in our case Ti is also seen to cluster in the same vecinity. Yet, its concentration is much higher towards the center of the supercell and not at the rich contiguous Nb-Ta regions. This also coincides with our Convex-Hull analysis in the previous section where Nb-Ti and Ta-Ti have both negative mixing entropy but they are overcome by the more stable Nb-Ta interaction.

\begin{figure}[!bt]
    \centering
    \includegraphics[width=\columnwidth]{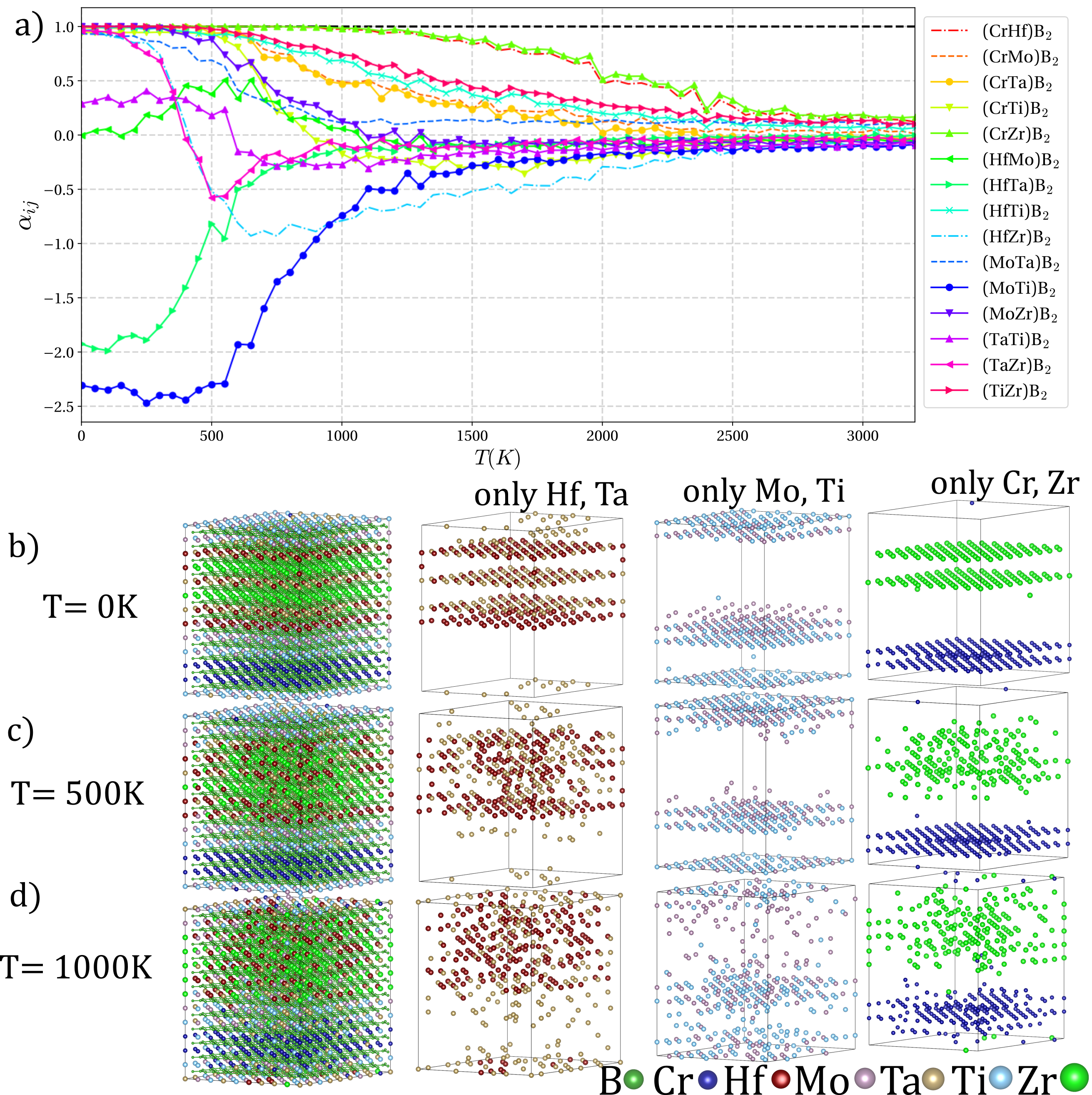}
    \caption{\textbf{SRO and Super-cell Graphs for the HEDB System.} (a) Illustrates SRO curves for the first nearest neighbor interactions within the active metal sublattice of the HEDB system. Panels (b), (c), and (d) depict the super-cell configuration at the last MC step for temperatures of 0, 500, and 1000 K, respectively. The first column visualizes all constituent atoms, with subsequent columns highlighting clustering atoms.}
    \label{fig:sro_hcp_full}
\end{figure}

Alternatively, the HEDB system shows a simpler ordering in its planar arrangement where atoms show clustering in different planes separated by inactive boron layers as shown in Figure \ref{fig:sro_hcp_full}. As in the previous case temperature slowly modifies the range of this clustering, going from 0K, 500K, and 1000K showing ordering change with temperature is shown in Figure \ref{fig:sro_hcp_full}a, b and c. In contrast to the BCC system, deviations from a random configuration remain significant well above 1,000 K. In this case, we observe the strong anti-clustering preference in both Zr and Cr and the pairing of low-energy pair interactions such as Hf-Ta and Mo-Ti clusters in the same plane at low temperatures. Additionally, it is important to note that Hf atoms can leak into Mo-Ti planes and there exists a Hf-Mo plane as well. The ordering in this periodic structure matches the pair energy profiles, where atoms share the same plane, or they delegate each other into another plane due to a better interaction available. We also declare that the superstructure geometry also dictates if the planes can fully cluster, that is, if the number of available atoms can fit into an integer number of planes. Here, 12 planes can accommodate the 6 equiatomic concentrations elements present in the composition.

We also notice the slow repulsion Cr poses in this system and how it diverges the least from its zero temperature lattice positions when increasing the temperature. We can track this phenomenon by observing the Cr pairs SROs, especially for Hf-Cr, and Cr-Zr, which retain a positive value well into the 2,000K. On the other hand, Zr loses its energetically stable clustering at higher temperatures, diverging into other planes where Hf and Ta have a higher presence. The pairs Hf-Ta and Mo-Ti cluster together, respectively, as shown in the SRO curves and the super-cell representations. Similarly, the overall less negative value for the Hf-Ta SRO is reflected in these atoms dispersing sooner when increasing temperature than their Mo-Ti counterpart.

\subsection{Investigating Ordering in Quaternaries and Quinaries in the BCC AlHfNbTaTiZr Alloy System}

Utilizing the Cluster Expansion (CE) technique, we conduct an ordering analysis on both quaternary and quinary subsystems. We generate two different supercells with a size of $8\times8\times8$ and $10\times7\times7$ conventional unit cells, accumulating to 1024 and 980 lattice sites, for the quaternary and quinary subsystems respectively. Randomly positioned constituent atoms are subjected to a Grand Canonical MC run for a period of 100,000 steps. To assure consistency in the final SRO value we select a burn-in period of 75,000 trials and we only average the SRO value of trials belonging to the last 25,000 steps.

To concisely present the findings across all examined subsystems, we compile and analyze the outcomes, to investigate the changes in entropy and Short-Range Order (SRO) variance across various systems and varying temperatures. The configuration entropy is calculated within the pair approximation of the Cluster Variation Method (CVM)\cite{tamm2015atomic, kikuchi1994cvm}:

\begin{equation}
\begin{split}
    S = k_B \Bigg( & \left( z - 1\right) \sum_i \left( c_i \ln{c_i} - c_i\right) \\
    & - \frac{z}{2} \sum_{ij} \left( P_{ij}^r \ln{P_{ij}^r} - P_{ij}^r\right) \\
    & + \left( \frac{z}{2} - 1\right) \Bigg)
\end{split}
\label{eqn:conf}
\end{equation}

\noindent where $z$ is the number of nearest neighbors, $c_i$ is the concentration of type $i$, and $P_{ij}^r$ is the probability of finding an $i,j$ pair, which in turn can be computed from the corresponding W-C parameter as calculated in Eq.~\ref{eqn:sro}. Additionally, we assess the variance of all SRO pair values at each temperature level as a measure of their deviation from the mean value at that temperature. This variance tends to decrease as temperatures rise, correlating with a convergence of SRO values towards zero and an overall trend towards full lattice disorder at $T \to \infty$.

\begin{figure*}
    \centering
    \includegraphics[width=\textwidth]{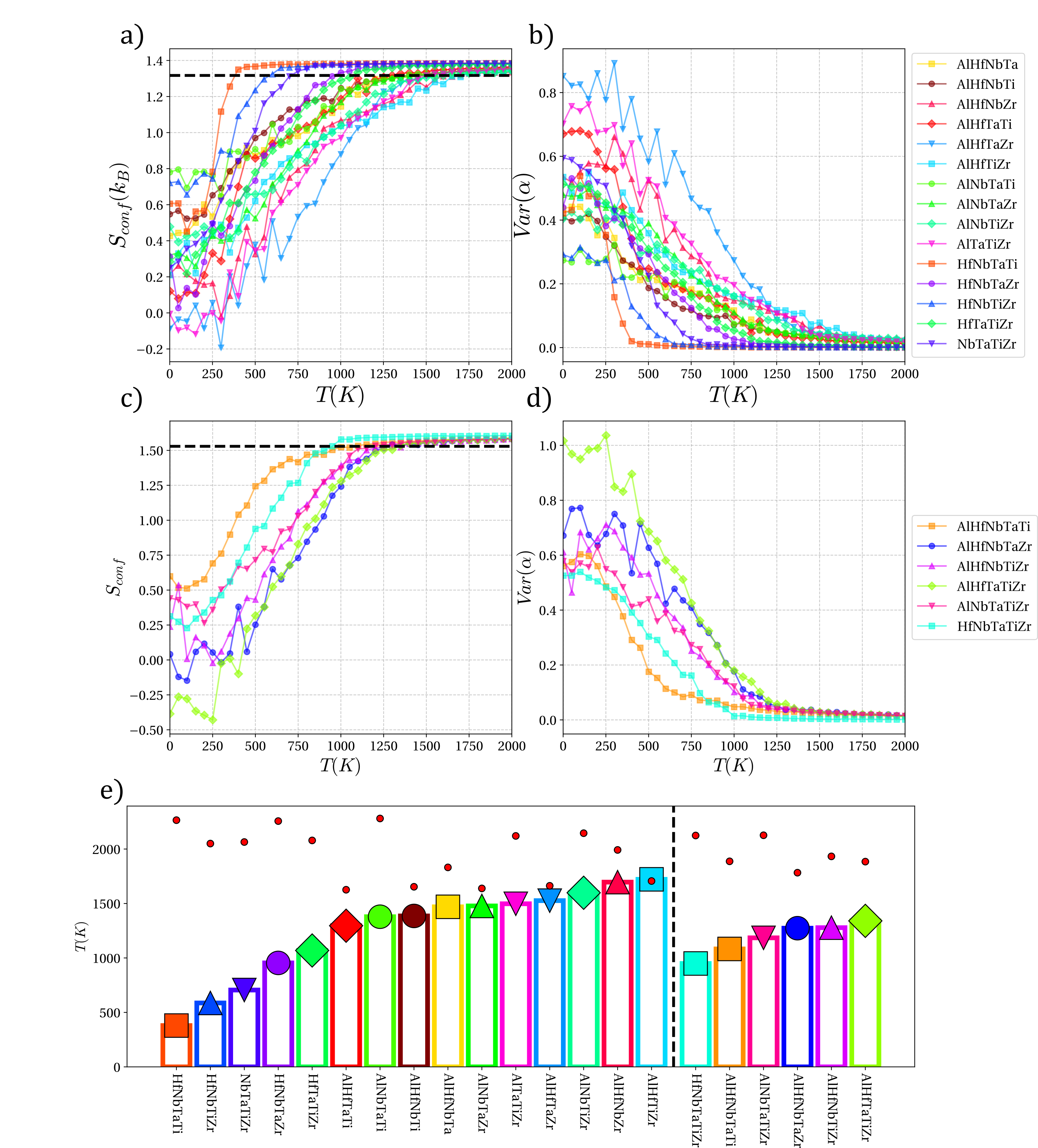}
    \caption{\textbf{Comparative Analysis of Entropy, SRO Variance, and Entropy Maxima in subsystems of the AlHfNbTaTiZr RHEA alloy space.} (a) and (c) display entropy as calculated using the Cluster Variation Method (CVM) for quaternary and quinary systems, respectively---dashed lines correspond to $0.95~S_{max}$. Panels (b) and (d) illustrate the variance in Short-Range Order (SRO) for these systems. Panel (e) presents the temperature at which alloys achieve 95\% of their maximum configurational entropy, with solidus temperatures depicted as small red dots. All panels consistently use the same labels for BCC subsystems, facilitating cross-comparison.}
    \label{fig:entropy}
\end{figure*}

In Figure \ref{fig:entropy}, quaternary subsystems are compared with other subsystems of equivalent complexity, highlighting key distinctions in their behavior at low temperatures, including their comparative degrees of ordering and the rates at which they lose this order. It is crucial to note that, at low temperatures, the configuration entropy is sometimes incorrectly predicted to be negative---a consequence of the Cluster Variation Method (CVM) pair approximation, as well as by the fact that the calculated configurations do not necessarily correspond to the equilibrium state due to lack of convergence.

Yet, once the alloy system begins to exhibit disorder, the configurational entropy calculation becomes accurate and converges to the theoretical maximum for a system of this size.

However, it is observed that quinary alloys, which exhibit high ordering at low temperatures---examples include AlHfTaTiZr, AlHfNbTiZr, and AlHfNbTaZr---are incorrectly predicted to have negative configurational entropy. Notably, these systems also achieve their maximum entropy later than others. This peak entropy coincides with the entropy curve reaching a threshold of 0.95 times the theoretical value for a completely disordered alloy.

In a similar vein, the two quaternary systems, AlHfTiZr and AlTaTiZr, which exhibit negative entropy at lower temperatures, also transition to a disordered state at relatively high temperatures. Thus, while the entropy values calculated using pair probabilities by the Cluster Variation Method (CVM) for highly ordered systems may not be reliable, the temperature at which these systems become disordered---reflected in their disorder temperature values---still occurs at relatively higher temperatures compared to similar subsystems.

Another commonality among the discussed systems is their inclusion of both Al and Zr. Remarkably, subsystems containing these elements consistently reach entropy values at the upper boundary of their entropy range at a higher temperature compared to similar systems. This implies that the presence of Al and Zr elevates the disorder temperature, attributed to the favorable interactions between these elements that persist at relatively high temperatures. A closer examination of the disorder trends within high entropy subsystems reveals that all Al-containing subsystems uniformly approach the upper limit of the disorder temperature spectrum. This phenomenon can be traced back to Al-Al interactions, which demonstrate probabilities lower than random even at temperatures significantly above the predicted solidus temperature. In essence, Al exhibits anti-clustering tendencies at lower temperatures, maintaining this behavior into temperature ranges where the bcc phase is presumably supplanted by a more stable liquid phase.

\subsection{Alloying Trends in Short Range Ordering of RHEAs}

\begin{figure*}
    \centering
    \includegraphics[width=\textwidth]{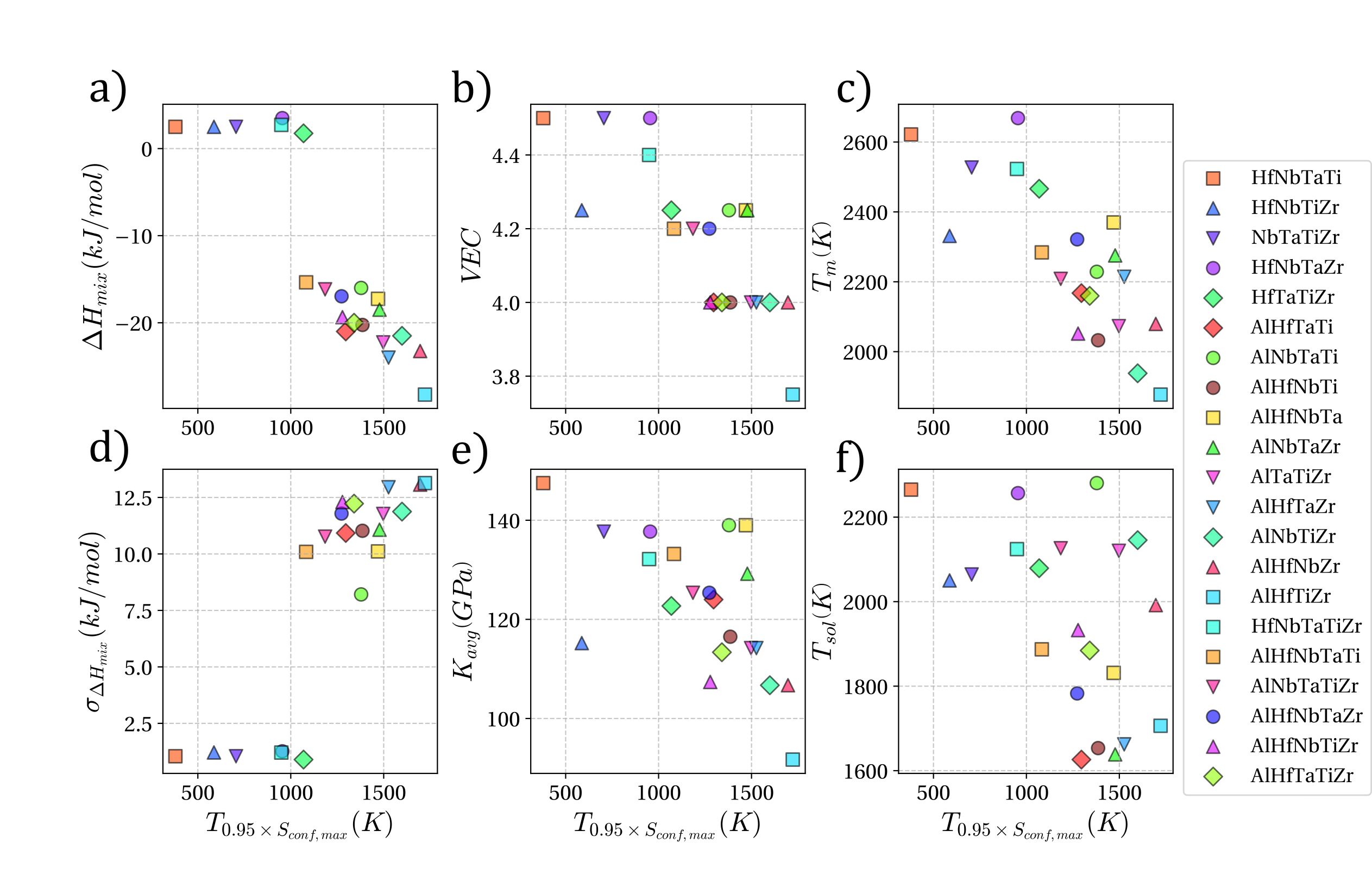}
    \caption{\textbf{Feature Correlation Analysis in HEA-ML Design for the AlHfNbTaTiZr RHEA alloy space.} Disorder temperature calculated against commonly utilized features: (a) Average Mixing Enthalpy, (b) Valence Electron Concentration (VEC), (c) Average Melting Temperature, (d) Standard Deviation of Mixing Enthalpy, (e) Average Bulk Modulus, and (f) CALPHAD Calculated Solidus Temperature. These comparisons aid in understanding the influence of each feature on the disorder temperature in high entropy alloys.}
    \label{fig:features}
\end{figure*}

In our final analysis, we integrate thermodynamically intensive calculations of disorder temperature with simpler thermodynamic features derived from unary and binary properties, as commonly adopted in the Machine Learning High Entropy Alloy (HEA) community to establish Hume-Rothery-inspired guidelines for HEA design \cite{zhou2019machine}. Figure \ref{fig:features} shows some of the trends observed between some low-order alloy metrics and the disordering temperature $T_{0.95 \times S_{conf,max}}$, defined as the temperature at which a given alloy system reaches 95 \% of the maximum possible configurational entropy---$S_{conf,max}=k_B \ln (4)$ for quaternaries, $S_{conf,max}=k_B \ln (5)$ for quinaries, etc.

Figure \ref{fig:features}~a) shows that a negative correlation exists between the average mixing enthalpy ($\Delta H_{mix}$) and the disordering temperature. The figure shows that there is no much correlation when $\Delta H_{mix}>0$, but the correlation strengthens for alloy systems with exothermic mixing enthalpies. When the average enthalpy of mixing is negative, it takes more thermal energy to disrupt chemical ordering. Fig. \ref{fig:features}~d)  shows that a positive correlation exists between $\sigma_{\Delta H}$ and the disordering temperature, indicating that the largest variance in the enthalpy of mixing between alloy pairs in a high order system is correlated to a higher disordering temperature. 

Figures \ref{fig:features}b) and e) explore the correlations between Valence Electron Concentration (VEC), average bulk modulus, and disorder temperature, revealing a slight negative trend, suggesting lower VEC values, which typically favor BCC formation, may elevate the disorder temperature, hinting at this feature's potential to predict chemically-ordering regions within BCC lattices.

Comparisons between melting temperature estimations—weighted average melting temperatures of an alloy system and CALPHAD calculations of the solidus temperature---from the TCHEA6 database---are presented in Figures \ref{fig:features}c) and f). An inverse relationship is observed between the averaged melting temperature and the disordering temperature. The correlation in this case may be related to the fact that alloys with lower average melting points contain Al and other elements with lower melting points---as discussed above, Al tends to stabilize chemical ordering in BCC RHEA systems. However, this trend is less pronounced with CALPHAD-calculated solidus temperatures, where a lower disorder temperature suggests a higher baseline for solidus temperatures, indicating the complex interplay between thermodynamic properties and disorder in HEAs.

\section{Conclusions}
\label{sec:conclusions}

This study has demonstrated the efficacy of a sophisticated framework that combines Cluster Expansion (CE) techniques with machine learning (ML), particularly leveraging M3GNet, to expedite and refine the thermodynamic analysis of compositionally complex systems. By starting from a relatively modest number of initial structures, this approach not only streamlines the exploration of the thermodynamic landscape of such systems but also achieves accuracy comparable to Density Functional Theory (DFT) calculations. This efficiency in generating and analyzing data has significantly simplified the fitting process of the CE model, allowing for the exploration of a broader array of configurations.

Our investigation reveals distinct interaction dynamics between BCC High Entropy Alloys (HEAs) and two-dimensional High Entropy Diboride (HEDB) systems. Notably, HEDB systems exhibit significant interaction parameters at triplets and long distances, facilitating the clustering of atoms within 2D planes. This suggests that, given adequate spatial freedom, atoms tend to organize into more energetically favorable configurations, clustering across different planes within the three-dimensional structure. 

Transitioning from simple binary systems to more intricate quaternary and quinary systems, although challenging, has provided valuable insights into the potential for atoms to cluster in more complex materials. The observed stability in simpler systems has illuminated pathways for understanding atomic grouping behaviors in more complex alloys, highlighting the potential of Short-Range Order (SRO) parameters as a tool for investigating phase stability and thermodynamics of short-range ordering.

Our study also integrates disordering temperature calculations with unary and binary thermodynamic alloy features to examine how the latter can be used in the design of HEAs that allow for chemical-ordering-based alloy optimization. The average mixing enthalpy (calculated from binary alloying pairs) as well as average melting point are inversely correlated with higher disordering temperatures. Likewise, lower VECs (corresponding to larger BCC stability) also result in higher disordering temperatures. While these design rules may not be generalizable beyond the specific system under study, they may be used to screen non-stoichiometric alloy chemistries in the AlHfNbTaTiZr RHEA alloy space.

Looking forward, the methodologies developed in this work pave the way for further exploration into the thermodynamics of compositionally complex systems. Future research will likely extend these analyses to a wider range of compositions, exploring the utility of SRO as a design parameter for enhancing material performance. Such investigations hold the promise of unlocking new chemistries where SRO can serve as a strategic lever for improving the properties of Refractory High Entropy Alloys and High Entropy Diborides.

In conclusion, this work not only advances our understanding of the thermodynamic behaviors of complex material systems but also showcases the transformative potential of integrating traditional computational techniques with modern machine-learning approaches. The ability to swiftly and precisely navigate both simple and complex systems heralds a new era for the innovative design of materials within the vast High Entropy domain, driving forward the discovery and optimization of novel materials for advanced applications.

\section*{Acknowledgements}
R.A. and G. V. acknowledge the support of QNRF under Project No. NPRP11S-1203-170056. D. S. acknowledges the support from NSF through Grant No. 1545403 (NRT-D$^3$EM). R. A. acknowledges support from NSF Grants No. 2119103 (DMREF) and 1905325 (DMREF). Research was also partly sponsored by the Army Research Laboratory and was accomplished under Cooperative Agreement Number W911NF-22-2-0106 (HTMDEC-BIRDSHOT program). The views and conclusions contained in this document are those of the authors and should not be interpreted as representing the official policies, either expressed or implied, of the Army Research Laboratory or the U.S. Government. The U.S. Government is authorized to reproduce and distribute reprints for Government Purposes notwithstanding any copyright notation herein. The authors acknowledge Texas A\&M High-Performance Research Computing (HPRC) for the computational resources used in this work.

\appendix

\end{document}